\RequirePackage{fix-cm}
\documentclass[twocolumn]{svjour3}          
\smartqed  
\usepackage{graphicx}
\usepackage{mathptmx}      
\usepackage{latexsym}
\usepackage{amsfonts} 
\usepackage{amssymb}
\usepackage{algorithmic}
\usepackage{algorithm}
\usepackage{amsmath}
\usepackage{multirow}
\usepackage[numbers]{natbib}

\begin{document}
\emergencystretch 3em

\title{A Novel Scalable Apache Spark Based Feature Extraction Approaches for Huge Protein Sequence and their Clustering Performance Analysis 
}

\titlerunning{60d-SPF and 6d-SCPSF}        

\author{Preeti Jha\textsuperscript{*} \and
        Aruna Tiwari \and
        Neha Bharill \and
        Milind Ratnaparkhe \and
        Om Prakash Patel \and
           Nilagiri Harshith   \and 
         Mukkamalla Mounika \and
     Neha Nagendra %
     }
 


\institute{P. Jha \at
              Indian Institute of Technology Indore \\
              \email{jha.preeti07@gmail.com}           
           \and
         	  A. Tiwari \at
              Indian Institute of Technology Indore \and
              N. Bharill \at
              Mahindra Ecole Centrale, Hyderabad  \and
              M. Ratnaparkhe \at
              ICAR-Indian Institute of Soybean Research Indore \and
              O. P. Patel \at
              Mahindra Ecole Centrale, Hyderabad  \and
                N. Harshith \at Mahindra Ecole Centrale, Hyderabad \and
              N. Nagendra \at
              Indian Institute of Technology Indore  \and
              M. Mounika \at
              Indian Institute of Technology Indore 
              }

\date{Received: date / Accepted: date}

\maketitle

\begin{abstract}
Genome sequencing projects are rapidly increasing the number of high-dimensional protein sequence datasets. Clustering a high-dimensional protein sequence dataset using traditional machine learning approaches poses many challenges. Many different feature extraction methods exist and are widely used. However, extracting features from millions of protein sequences becomes impractical because they are not scalable with current algorithms. Therefore, there is a need for an efficient feature extraction approach that extracts significant features. We have proposed two scalable feature extraction approaches for extracting features from huge protein sequences using Apache Spark, which are termed 60d-SPF (60-dimensional Scalable Protein Feature) and 6d-SCPSF (6-dimensional Scalable Co-occurrence-based Probability-Specific Feature). The proposed 60d-SPF and 6d-SCPSF approaches capture the statistical properties of amino acids to create a fixed-length numeric feature vector that represents each protein sequence in terms of 60-dimensional and 6-dimensional features, respectively. The preprocessed huge protein sequences are used as an input in two clustering algorithms, i.e., Scalable Random Sampling with Iterative Optimization Fuzzy c-Means (SRSIO-FCM) and Scalable Literal Fuzzy C-Means (SLFCM) for clustering. We have conducted extensive experiments on various soybean protein datasets to demonstrate the effectiveness of the proposed feature extraction methods, 60d-SPF, 6d-SCPSF, and existing feature extraction methods on SRSIO-FCM and SLFCM clustering algorithms. The reported results in terms of the Silhouette index and the Davies-Bouldin index show that the proposed 60d-SPF extraction method on SRSIO-FCM and SLFCM clustering algorithms achieves significantly better results than the proposed 6d-SCPSF and existing feature extraction approaches. \end{abstract}

\keywords{Apache Spark Cluster \and Big Data  \and Feature Extraction  \and Fuzzy Clustering  \and Huge Protein Sequences  \and Scalable Algorithms}

\section{Introduction} \label{sec1}
Bioinformatics has become a major field of research and is gaining the constant attention of biologists and computer scientists. The main goal of bioinformatics is to collect and manage biological data and to develop data mining techniques that are useful for the analysis and understanding of biological processes. With the fast improvement of Next-Generation Sequencing (NGS) innovation, a huge number of genomic datasets have been created, representing a great challenge to customary bioinformatics tools \cite{p9, 2020scalable}. The surprising development of protein sequences is a problem in managing these sequences to a large extent. The number of unique sequences in the protein database together now surpasses a million. A protein family contains a large number of sequences that are progressively related \cite{p1}.  The clustering of protein sequences aims to provide meaningful partitioning from a huge protein dataset. The protein sequences are arranged into clusters based on their similarity in protein sequences \cite{2020sequence}. 
 Clustering protein sequences predicted from sequencing reads can impressively reduce the excess of sequence sets and the expense of downstream analysis and storage \cite{p2,2020ml}. Many researchers have worked on the K-means clustering algorithm to create high-quality sequence clusters \cite{p3,p4}. However, the K-means algorithm calculates the distance between the data samples with exact precision. When the distance function is not properly defined, the K-means method may fail to disclose the sequence-to-structure relationship. As a result, many clusters have poor protein sequence matching. Hence, due to poor protein sequence clustering, the fuzzy concept has emerged as a new research frontier \cite{jha,jha2021apache,jha2021scalable}.

A data sample in fuzzy clustering might belong to many clusters with varying degrees of membership.
The Fuzzy c-Means (FCM) technique was introduced by Bezdek \cite{r5} and employs iterative optimization to minimize an objective function using a similarity measure on feature space.  Many researchers have worked with the FCM algorithm for protein sequence clustering \cite{p5,p6,p7}.
Due to the rapid evolution of clustering algorithms, these techniques have gained less recognition for dealing with Big Data challenges.
This is because these methods are not scalable \cite{r20}.
Big Data environments are required to build scalable methods for dealing with enormous amounts of data generated from many sources \cite{r21}. 
To handle large amounts of data, several fuzzy clustering methods, such as the FCM extension, also known as random sampling plus extension Fuzzy c-Means (rseFCM), are employed. Still, the cluster overlapping is an issue with rseFCM. The overlap is eliminated by Random Sampling with Iterative Optimization Fuzzy c-Means (RSIO-FCM) \cite{srsio}. However, RSIO-FCM, suffers from a rapid rise in multiple iterations during clustering.  To address the shortcomings of RSIO-FCM, a scalable incremental fuzzy clustering technique known as SRSIO-FCM  was designed \cite{srsio}.  The SRSIO-FCM inherently makes use of the Scalable Literal Fuzzy c-Means (SLFCM) algorithm during the clustering process. In the literature survey, it is found that the SRSIO-FCM algorithm generates clusters of better quality in comparison with the RSIO-FCM method.
 Therefore, in this work, we have used the SRSIO-FCM and SLFCM algorithms to cluster huge protein sequences. The SRSIO-FCM method is implemented in Apache Spark \cite{srsio}. It splits large amounts of data into smaller subsets. For the first subset, the  membership degree belonging to the cluster centers is generated by applying the SLFCM algorithm \cite{srsio}. The first subset cluster centers are assigned to the second subset for clustering. But unlike RSIO-FCM, it does not use these cluster centers as an input for the clustering of the third subset. Instead, it combines the membership information of the first and second subset to compute the new cluster centers. The membership matrices from the first two subsets are merged, and the cluster centers are fed into the third subset. This procedure is repeated for the next grouping of all subgroups. In RSIO-FCM, for clustering of the next subset, the cluster center of one subset is fed as an input next subset. Hence, it will result in slow convergence by taking the higher number of iterations for that subset. This problem has been overcome in the SRSIO-FCM, because this approach uses the cluster centers obtained with the combined membership matrices of all the processed subsets for the clustering of the current subset. 

Before applying any machine learning algorithm to protein sequences for analysis and modeling of huge sequences, it is important to encode the protein sequences into feature vectors. Protein sequences are made up of characters from the 20-letter amino acid alphabets $\sum$=\{A, C, D, E, F, G, H, I, K, L, M, N, P, Q, R, S, T, V, W, Y\}. The amino acids of a sequence can be concatenated in any order, and the protein sequences can be of any length.  The high dimensionality of protein data
presents many critical challenges while applying any machine learning methods \cite{vipsita2013two}. Therefore, representation of protein sequences in terms of feature
vectors is an important problem to be addressed. For the correct categorization of protein sequences, a suitable input representation (feature extraction) is required. Wang
et al. \cite{wang2001new} developed a new feature extraction approach that attempts to capture both global and local similarity of sequences.
Local similarity refers to frequently occurring sub-strings in the sequences, whereas global similarity refers to the overall similarity among numerous sequences.  It has considered the 2-gram method as discussed by Wu
et al. \cite{wu1992protein} to compute the global similarity of  sequences and adopted a method called 6-letter exchange groups to represent a sequence \cite{dayhoff1978model}. Then it uses a sequence mining tool to compute the local similarity. Das et al. \cite{f21} developed a mapping sequence to feature vector using a numerical representation of codons targeted to amino acids for alignment-free sequence analysis. Bandyopadhyay \cite{bandyopadhyay2005efficient} proposes another feature extraction strategy that employs a 1-gram algorithm for feature encoding.
The feature size consists of 20 amino acids.
The extracted features are  taken into account the probability of occurrences of the amino acids in the various positions of the sequences.  G Mansoori et al. \cite{g2008generating} suggested a novel feature extraction approach.
To extract important characteristics from protein sequences, the occurrences of six exchange groups \cite{dayhoff1978model} in each sequence are counted. 
Chou \cite{chou1} proposed two types of models widely used to illustrate protein models: sequential models and descriptive models, although both the models suffer certain disadvantages. The sequential model fails when the query protein has no substantial sequence resemblance to any known protein.
As a result, Chou offers a distinct model known as the pseudo-amino acid composition model \cite{chou1,chou3}. Gupta et al. \cite{60d} used the general version of Chou \cite{chou2} pseudo amino acid composition, which is a sixty-dimensional numerical feature vector of protein sequences, to develop an alignment-free approach for finding similarity across protein sequences. Bharill et al. \cite{6d-c} developed an approach to extract six-dimensional numerical feature vectors from a protein sequence.  Likewise, many feature extraction techniques \cite{f21, wang2001new,  bandyopadhyay2005efficient, mansoori2009protein, g2008generating, 6d-c, 60d} have been introduced in the past, but, none of them is scalable. A scalable approach for selecting statistically relevant characteristics from a large sequence is required. 

Many Big Data processing frameworks exist to make the approaches scalable for processing huge protein data. In this work, to make the proposed feature extraction approaches scalable, we have used the Apache Spark Big Data processing framework \cite{r22,r23}. Apache Spark is a general-purpose distributed processing system used for Big Data framework built on top of the Hadoop Distributed File System (HDFS). It utilizes in-memory computation and optimized query execution for fast analytic queries against data of any size \cite{spark,rdd,spark-bigdata}. We propose two scalable preprocessing algorithms for feature extraction of huge protein sequences using the Apache Spark framework; a 60-dimensional Scalable Protein Feature extraction approach named 60d-SPF and a 6-dimensional Scalable Co-occurrence based Probability Specific Feature extraction approach named 6d-SCPSF. After extracting fixed length numeric feature vectors from huge protein sequences using these approaches, the numeric feature vectors are passed as input to the scalable clustering algorithms, i.e., SRSIO-FCM and SLFCM, to create clusters for huge protein datasets. Finally, the validation measures are used to validate the results obtained from the SRSIO-FCM and SLFCM algorithms, which helps to analyze the performance of the proposed 60d-SPF and 6d-SCPSF feature extraction methods. Additionally, we have compared the clustering results of the proposed 60d-SPF and 6d-SCPSF feature extraction methods with other existing feature extraction methods \cite{bandyopadhyay2005efficient, mansoori2009protein}. 

This paper is broadly standardized as follows: Section \ref{sec2} describes feature extraction methods for protein sequences. The Apache Spark framework is discussed in Section \ref{apache}.  In Section \ref{sec3}, the implementation of the proposed 60d-SPF and 6d-SCPSF extraction algorithms using Apache Spark is explained in detail. In Section \ref{sec4},  we have reported the results of the experimental study obtained from the proposed 60d-SPF and 6d-SCPSF methods, as well as other existing methods when applied to the SRSIO-FCM and SLFCM algorithms for clustering of huge protein sequences. The results are reported in terms of the Silhouette index and Davies Bouldin index. Finally, the conclusions of our work are drawn in Section \ref{sec5}.

\section{Preliminaries}
\label{sec2}
 In this section, we present a detailed description of the existing feature extraction techniques applied to protein sequences that form the basis for the discussion of the proposed feature extraction approaches.

A 60-dimensional feature vector corresponding to each protein sequence is formed by considering three properties of each amino acid from 20 amino acids \cite{chou2,chou1,chou3,chou2005using,60d}.  The numerical construction of the vector is based on the following:
\begin{enumerate}
\item Amino acid count:
The twenty amino acids for protein sequences are $\sum$=\{A, C, D, E, F, G, H, I, K, L, M, N, P, Q, R, S, T, V, W, Y\}.  The total length of 20 amino acids defined as $l_A, l_C, l_D, l_E, l_F, l_G, l_H, l_I, l_K, l_L, l_M, l_N, l_P, l_Q, l_R, l_S, l_T, l_V, l_W, \\ l_Y$. 
The total number of amino acids in a protein sequence is determined by the length of the sequence. This criterion is insufficient since sequences with the same number of amino acids in various places differ substantially. As a result, in a 60-dimensional vector, it contributes to the first 20 values \cite{60d}.
    \item Total distance: 
   The sum of the distances between each amino acid and the first amino acid in the protein sequence is the total distance.
The other 20 values in the 60-dimensional vectors are influenced by these 20 values.
However, for identical protein sequences, this parameter may appear to be the same. 
     The total distance of $T_i$ is defined as follows:
\begin{equation}
\label{Ti}
T_i = \sum _{j=1}^{l_i} T_j 
\end{equation}
where, $i =A, C, D, E, F, G, H, I, K, L, M, N, P, Q, R, S, T, V, \\W, Y$ ; $t_j$ is the distance from the first amino acid to the $j^{th}$ amino acid of $i$ in the protein sequence. The other twenty feature vectors of the total distances denoted as $T_A, T_C, T_D, T_E, T_F, T_G, T_H, T_I, T_K, T_L,T_M, T_N, T_P, T_Q, T_R, T_S, \\T_T, T_V, T_W, T_Y$.
    
    \item Distribution: Two parameters, as seen above, are unable to accurately differentiate the similarity/dissimilarity of the two sequences.
As a result, the third parameter, $D_i$, represents the distribution of 20 amino acids throughout the protein sequence.
Even if two protein sequences have the same content and total distance of 20 amino acids, their amino acid distribution are different.
As a result, one-third of the 60-dimensional vector is made up of the 20 amino acid distribution.
Each amino acid distribution is determined as follows: 
    \begin{equation}
\label{Di}
D_i = \sum _{j=1}^{l_i} \frac{(t_j - d_{i})^2} {l_i} 
\end{equation}
where, $i =A, C, D, E, F, G, H, I, K, L, M, N, P, Q, R, S, T, V, \\ W, Y$ ; $t_j$ is the distance from the first amino acid to the $j^{th}$ amino acid of $i$ in the protein sequence and 
$d_i= \frac{T_i}{l_i}$. We can then say that all these three characteristics make up the 60-dimensional vector that characterizes the protein sequence.  The 60-dimensional equal length vectors were constructed for unequal length protein sequences. So, the feature vector, which contains 60-dimensional data, is given as follows:\\
 $< l_A, T_A,  D_A, l_C, T_C, D_C, l_D, T_D, D_D, l_E, T_E, D_E, l_F, T_F, D_F,\\ l_G, T_G, D_G, l_H, T_H, D_H, l_I, T_I,  D_I, l_K, T_K,  D_K, l_L, T_L, D_L, l_M, \\T_M, D_M,  l_N, T_N, D_N, l_P, T_P, D_P, l_Q, T_Q, D_Q, l_R,  T_R, D_R, l_S, T_S, \\ D_S, l_T, T_T, D_T, l_V, T_V, D_V, l_Y, T_Y,  D_Y >$

\end{enumerate}
The 60-dimensional approach extracts 60 feature vectors from a protein sequence. The detailed description of the six-dimensional feature extraction method are presented next.

A 6-dimensional feature extraction approach corresponding to each protein sequence is formed by considering three stages \cite{6d-c,dayhoff1978model,dayhoff1972model}. The 6-dimensional feature extraction approach captures the protein sequences statistical properties together with the amino acid position information to generate a vector of fixed-length numerical features for each protein sequence.
The Co-occurrence-based Probability-Specific Feature (CPSF) extraction approach is a 6-dimensional feature extraction approach that extracts features from the protein dataset in three steps: protein sequence encoding (PSE), global similarity measure (GSM), and local similarity measure (LSM). The first step encodes protein sequences, representing each protein sequence as exchange groups. In the second stage, the overall GSM is calculated, considering the probability that each amino acid appears at a particular position within the total number of protein sequences present in a specific species. In the third stage, the LSM calculates each amino acid weight concerning each protein sequence, taking into account the GSM. In this way, the 6-dimensional feature extraction approach represents each protein sequence by a fixed-length numerical vector consisting of only six dimensional numeric feature vectors. 
\subsubsection{Stage One: Protein Sequence Encoding (PSE)}
\label{stage1}
In the first step of the CPSF approach \cite{6d-c}, each protein sequence is encoded and presented in terms of six exchange groups. According to Dayhoff and Schwartz \cite{dayhoff1978model}, the amino acids in the protein sequence belong to six exchange groups. This is because, within each exchange group, there is a high evolutionary similarity between these amino acids. Exchange groups are efficient amino acid equivalent classes, formally represented by $\{e_{1}, e_{2}, e_{3}, e_{4}, e_{5}, e_{6}\}$, where $e_{1}$=$\{H, R, K\}$, $e_{2}$=$\{D, E, N, Q\}$, $e_{3}$=$\{C\}$, $e_{4}$=$\{S, T, P, A, G\}$, $e_{5}$=$\{M, I, L, V\}$ and $e_{6}$=$\{F,Y,W\}$ \cite{wang2001new}. The protein sequences belong to different species, and within each species, the protein sequences share some structural similarities. Once the sequence of proteins is encoded by the exchange groups, the total similarity between the encoded exchange groups is calculated. A detailed description of the subsequent stages is presented next.

\subsubsection{Stage Two: Global Similarity Measure (GSM)}
\label{stage2}
In the second step of the CPSF method, we calculate the GSM by estimating the instance probability of all exchange groups at each position relative to the total number of protein sequences in the species. The GSM is calculated as follows: 
\begin{equation}\label{eq:E9}
(Probability)_{ij}=(Instance)_{ij}/\eta
\end{equation}
Where $(Probability)_{ij}$ denotes the probability of instance of the $i^{th}$ exchange group at $j^{th}$ position, $(instance)_{ij}$ represents the frequency at which the $i^{th}$ exchange group appear at $j^{th}$ position and $\eta$ represents the total number of sequences in a particular species. 
After that, LSM is calculated, which determines the specific weight of each exchange group's position. A detailed description of it is given in the subsequent section.
\subsubsection{Stage Three: Local Similarity Measure (LSM)}
\label{stage3}
In the third stage of the CPSF approach \cite{6d-c}, the LSM is calculated, which determines the location-specific weight of each exchange group within the sequence considering the weight factors. These weight factors ultimately represent the numeric feature vectors for each protein sequence. The weight of each exchange group is calculated as follows:
\begin{equation}\label{eq:E10}
(Weight)^{SEQ_{k}}_{i}=\sum_{j=1}^{j^{\prime}}(Probability)_{ij}\times (PW)_{ij}^{SEQ_{k}}
\end{equation}
where $(Weight)^{SEQ_{k}}_{i}$ represents the weight of $i^{th}$ exchange group corresponding to the $k^{th}$ protein sequence, $(Probability)_{ij}$ denotes the probability of occurrence of the $i^{th}$ exchange group at $j^{th}$ position and $(PW)_{ij}^{SEQ_{k}}$ is the positional weight assigned to the $i^{th}$ exchange group based on the presence of $k^{th}$ protein sequence at $j^{th}$ position. 

The CPSF approach extracts a numeric feature vector from the protein sequence. The extracted numeric feature vector consists of only six dimensions.

We have adopted the 60-dimensional approach and propose the scalable feature extraction approach named 60d-SPF which has been discussed in detail in Section \ref{SPF}.
Furthermore, we have proposed and designed a scalable version of the CPSF feature extraction method named 6d-SCPSF. A detailed description of the 6d-SCPSF approach has been given in Section \ref{SCPSF}. Before introducing the proposed algorithms, we give the details of a well-known framework for Big Data processing in Section \ref{apache}.

\section{Big Data Framework: Apache Spark}
\label{apache}
\begin{figure}[t]
  \includegraphics[width=0.50\textwidth]{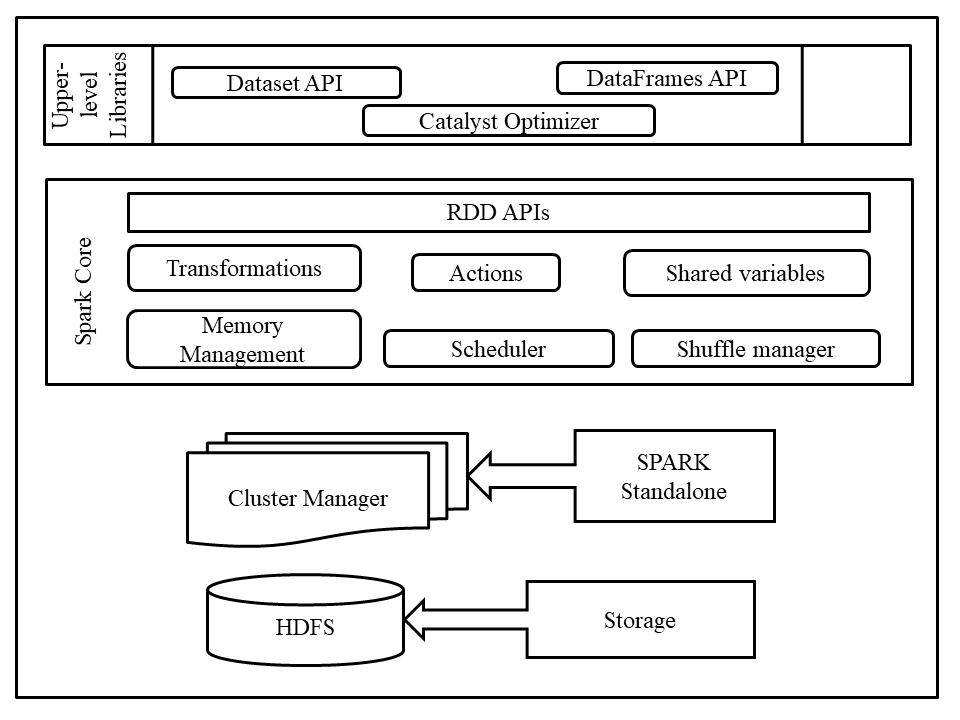}
\caption{Apache Spark cluster stack.}
\label{spark-stack}      
\end{figure}

The proposed 6d-SCPSF and 60d-SPF extraction algorithms can be scalable using Apache Spark clusters to handle large-scale protein data. To speed up implementation of the feature extraction approach in this paper, we choose Apache Spark to implement the proposed algorithms. Apache Spark consists of several main components, including the Spark core and higher-level libraries. Figure \ref{spark-stack} shows the Apache Spark clusters stack used in our experiment, and the details for the same are explained below. 

\noindent \textbf{\textit{Spark core}}:
The Spark core runs on different cluster managers and can access data from any Hadoop data source. It provides a simple programming interface for large-scale processing datasets, Resilient Distributed Dataset (RDD). Spark Core is embedded in Scala, but it comes with APIs in Scala, Java, Python, and R. Python in our case. Besides, the Spark core provides a key function for in-memory cluster computing, including memory management, job scheduling, data shuffling, and error recovery \cite{apache2016}.

\noindent \textbf{\textit{Upper-level libraries}}:
Spark SQL has been created to manage various workloads on the Spark core. 

\noindent \textbf{\textit{Cluster managers and data source}}:
A cluster manager is used to obtain cluster resources to run a job. The Spark Engine works with the built-in Spark cluster manager (i.e., standalone). The cluster manager manages resource sharing between Spark applications. On the other hand, Spark can access data from the Hadoop Distributed File System (HDFS) \cite{hdfs}. 

\noindent \textbf{\textit{Resilient Distributed Datasets}}:
Spark Core is built on the RDD abstraction. An RDD is a read-only distribution of records. RDDs provide a false tolerance, a parallel data structure that allows users to explicitly store data on disk or memory, control its partitions, and manipulate it using a full set of operators. This enables us to efficiently distribute the data in the calculation of the necessary requirements for different workloads. An RDD can be created from external data sources or from other RDDs. 

\begin{figure}[ht] 
  \includegraphics[width=0.48\textwidth]{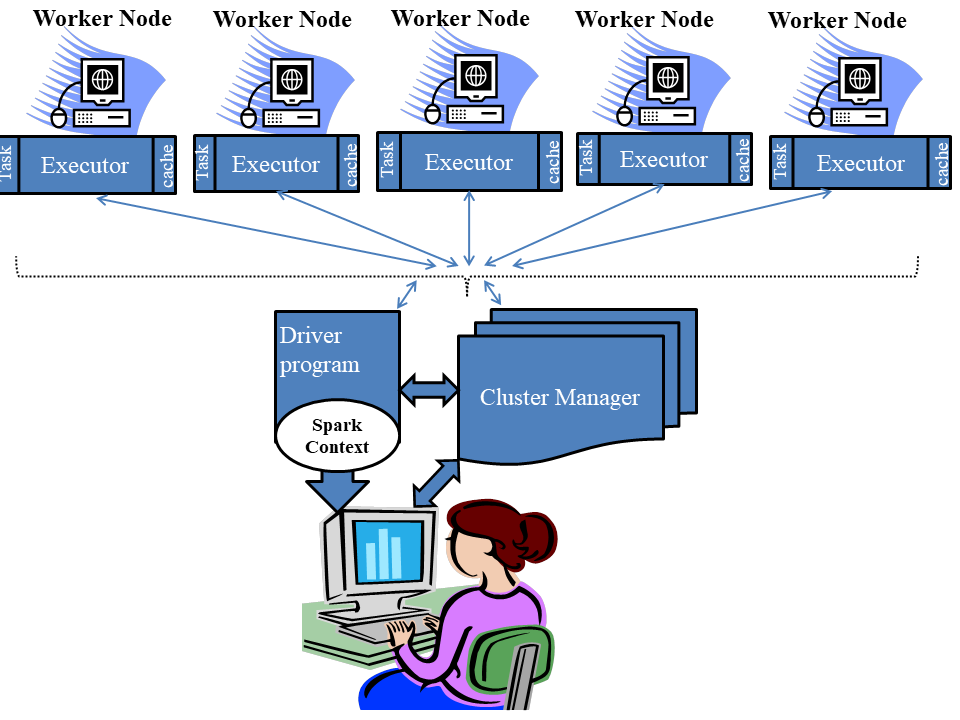}
\caption{Apache Spark cluster Application.}
\label{spark-cluster}      
\end{figure}

\noindent \textbf{\textit{Spark-application}}: The Spark application execution consists of five main units, a controller program, a cluster administrator, workers, executors, and tasks. The Apache Spark clusters application is shown in Figure \ref{spark-cluster}. A driver program is an application that uses Spark as a library and defines the high-level flow of control for the target calculation. While a worker provides CPU, memory, and storage resources to a Spark application, an executor is a Java Virtual Machine (JVM) process that Spark creates on each worker for that application. A job is a set of calculations that the Spark controller performs on a cluster to get results in the program. A Spark application can start multiple jobs. Spark divides the work into steps of a directed acyclic chart (DAG), where each phase is a collection of tasks. A task is the smallest unit of work that a spark sends to an executor. The main entry point for spark functions is a spark context, through which the driver program uses Spark. A Spark context represents a connection to a computing cluster.

The proposed algorithms were implemented using the Apache Spark clusters framework, where Hadoop is used as a data storage. The proposed 60d-SPF and 6d-SCPSF extraction methods are explained in Section \ref{sec3}. 

\section{Proposed Work}
\label{sec3}
This section describes the proposed scalable algorithms implemented on the Apache Spark cluster. To propose a scalable protein preprocessing algorithm, we followed the  pseudo-amino acid composition approach discussed in \cite{60d} and proposed the design of a scalable feature extraction approach named 60d-SPF. The proposed 60d-SPF approach, when applied to huge protein data, extracts 60-dimensional numeric feature vectors. In addition to this, we propose the design of another scalable feature extraction named 6d-SCPSF, which is inspired by the CPSF approach \cite{6d-c}. To make the proposed feature extraction approaches scalable, we implemented them on the Apache Spark cluster. The output obtained from the scalable feature extraction approaches is passed as an input to the SRSIO-FCM and SLFCM \cite{srsio} clustering algorithms to form clusters of the huge protein sequence data. A detailed description of the proposed scalable feature extraction approaches is presented next.

\subsection{Scalable 60-dimensional Feature Extraction Approach}
\label{SPF}
\begin{figure}[ht] 
  \includegraphics[width=0.5\textwidth]{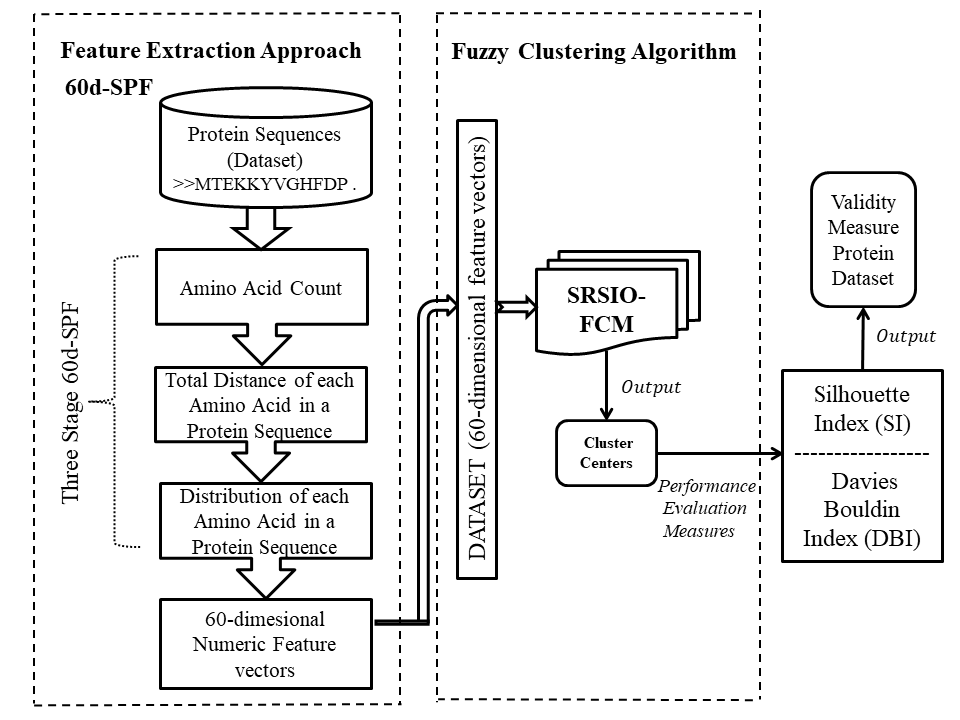}
\caption{Workflow of 60d-SPF Architecture.}
\label{60d-architecture}      
\end{figure}

To represent all protein sequences in terms of 60-dimensional numeric feature vectors, the proposed scalable protein feature extraction technique is implemented using the Apache Spark framework.  The architecture of the proposed 60d-SPF extraction approach is shown in Figure \ref{60d-architecture}. 
Algorithm \ref{Protein-algo} summarizes the steps of the proposed 60d-SPF approach. The input given is a raw protein dataset containing 20 amino acids $\sum$=\{A, C, D, E, F, G, H, I, K, L, M, N, P, Q, R, S, T, V, W, Y\}. The output is a feature vector of the given input dataset, which is a file containing 60-dimensional numeric feature vectors.

\begin{algorithm}[t]
  \caption{60d-SPF }
\label{Protein-algo}
\begin{algorithmic}[1]
\REQUIRE Raw protein data : $raw\_Protein.txt$
 \ENSURE Processed protein data : $ proc\_Protein.txt$
\STATE \textbf{Read} the protein sequences from the fasta file ($raw\_Protein.txt$) and parallelize them with the help of Spark RDD.
\STATE \textbf{Map} each sequence in RDD to $Protein\_Preprocess()$ algorithm for that particular sequence.
 \STATE \textbf{Save} feature vectors to text file: $proc\_Protein.txt$.
\end{algorithmic}
\end{algorithm}
 
 In Line 1 of Algorithm \ref{Protein-algo} , the data is read into an RDD from the Hadoop.  Then a map function is used to accomplish the preprocessing task. In Line 2 of Algorithm \ref{Protein-algo}, the map function calls the Algorithm \ref{Protein-sub}: $Protein\_Preprocess$,  where each sequence is mapped into RDD and then applied $Protein\_Preprocess$ algorithm for that sequence. Finally, the obtained RDD after map function is saved using Line 3 of Algorithm \ref{Protein-algo}. The proposed scalable feature extraction approaches extracts features of protein sequences in three sets of the numerical parameter. The first parameter is a calculation of the length of sequences. The second parameter is the total distances of each amino base to the first amino. The third parameter is the variance of distance for each amino acid \cite{60d}. Each set of numerical parameters is not sufficient to denote a specific protein sequence. Thus, a combination of all the three sets of numerical parameters will represent each protein sequence with a 60-dimensional numeric feature vector.  The Algorithm \ref{Protein-sub} discusses the $Protein\_Preprocess$ algorithm, which is called by 60d-SPF given in Algorithm \ref{Protein-algo}. 
 
\begin{algorithm}[t]
\caption{$Protein\_Preprocess$}
\label{Protein-sub}
\begin{algorithmic}[1]
\REQUIRE Grid of values; $z$: [A, C, D, E, F, G, H, I, K, L, M, N, P, Q, R, S, T, V, W, Y]
  \ENSURE $l_i, T_i, D_i$ 
   
 \STATE Let $i$ denote amino $\sum$=\{A, C, D, E, F, G, H, I, K, L, M, N, P, Q, R, S, T, V, W, Y\}.
\FOR{ $x$ in $z$}
 \IF{$x$ is $i$}
\STATE Increase the count of amino $i$ i.e., $l_i++$; 
\ENDIF
\ENDFOR
\STATE \textbf{Calculate} total distance $T_i$ using Eq. \ref{Ti}.
\STATE \textbf{Calculate} the variance of distance $D_i$ using Eq. \ref{Di}.
\end{algorithmic}
\end{algorithm}

The proposed scalable protein preprocessing algorithm (60d-SPF) has the significant characteristic that it takes raw protein sequences as input and produces 60-dimensional numeric feature vectors as an output which is feed as an input to SRSIO-FCM and SLFCM clustering algorithms to produce output in terms of clusters of protein sequences.
The working of the $Protein\_Preprocess$ algorithm is shown in various steps by using a toy illustration presented next.
\\

\begin{figure}[ht]
    \centering
 \includegraphics[width=0.50\textwidth]{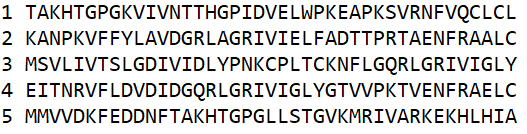}   
 \caption{Example of protein sequences.}
    \label{sequences}
\end{figure}

\noindent  \textbf{A Toy Illustration for 60d-SPF Approach}\\
Here, we give a toy illustration of the five protein sequences as shown in Fig. \ref{sequences} to explain the $Protein\_Preprocess$ algorithm.  The toy problem is solved using the following steps.\\
\noindent \textbf{Step-1:} \textit{Calculation of length of sequence}

\begin{figure*}[!t]\centering\setlength\fboxsep{0.2pt}\setlength\fboxrule{0.25pt}\fbox{\includegraphics[width=7.2in, height=1.3cm]{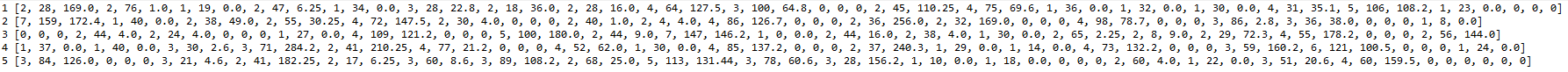}}\caption{Preprocessed result using proposed 60d-SPF extraction method for the protein sequences given in Fig.  \ref{sequences}.}\label{sequence-result}\end{figure*}

\noindent The $Protein\_Preprocess$ algorithm (given in Algorithm \ref{Protein-sub}) working is being described using a first step in which the length of each amino acid sequence is calculated as shown in Line 1-6. Here an illustration is presented by considering an example of five sequences are shown in Fig. \ref{sequences}. The output obtained after preprocessing of these five sequences are shown in Fig. \ref{sequence-result}, where the first column is the sequence number and the other information presents in the form of 60-dimensional numeric feature vectors are as follows: \\ $< l_A, T_A,  D_A, l_C, T_C, D_C, l_D, T_D, D_D, l_E, T_E, D_E, l_F, T_F, D_F, l_G, \\T_G, D_G, l_H, T_H, D_H, l_I, T_I,  D_I, l_K, T_K,  D_K, l_L, T_L, D_L, l_M,  T_M,\\ D_M,  l_N, T_N, D_N, l_P, T_P, D_P, l_Q, T_Q, D_Q, l_R,  T_R, D_R, l_S, T_S, D_S, \\l_T, T_T, D_T, l_V, T_V, D_V, l_Y, T_Y,  D_Y >$

The sequence1 of Fig. \ref{sequences} is represented in terms of feature vectors are $[2,2,1,2,1,3,2,2,2,4,3,0,2,4,1,1,1,4,4,4,\\1,0]$, respectively. Likewise, the value of the other four sequences are shown in Fig. \ref{sequence-result}. 

\noindent \textbf{Step-2:} \textit{Total distances of each amino acid to the first amino acid}

\noindent The second numerical parameter, i.e., the total distances of each amino acid-base to the first amino acid $T_i$, which is calculated using Eq. \ref{Ti} in Line 7 of $Protein\_Preprocess$ algorithm. Add the position values corresponding to each amino acid as shown in Fig. \ref{sequences}, and amino acid $G$ appears at 5, 7, and 16 (index starts from $0$). So, the value of $T_G$ is calculated as follows:
\begin{center}
   $T_G$= $5+7+16$ = $28$
\end{center}
 Likewise, the value of  $T_A, T_C, T_D, T_E, T_F, T_H, T_I, T_K, T_L,T_M, T_N, \\ T_P, T_Q, T_R, T_S, T_T, T_V, T_W, T_Y$ for sequence1 and the total distances of each amino acid to the first amino acid for all other four sequences are shown in Fig. \ref{sequence-result}.
 
\noindent \textbf{Step-3:} \textit{Variance of distance for each amino acid}

\noindent The third numerical parameter $D_i$ is the variance of distance for each amino acid, which is calculated using Eq. \ref{Di} in Line 8 of $Protein\_Preprocess$ algorithm. The first step is to compute $d_i$ using $d_i= \frac{T_i}{l_i}$. The value of $d_G$ in sequence1 is calculated as follows:
\begin{center}
   $d_G= \frac{T_G}{l_G}$= $\frac{31}{3}$ =$10.33$  
\end{center}
The second step is to compute the variance of distance for each amino acid-base, i.e., $D_i$, for example, the value of $D_G$ in sequence1 is $22.88$, calculations of $D_G$ is as follows:
\begin{center}
 $D_G$ = $\frac{[(6-10.33)^2+(8-10.33)^2+(17-10.33)^2]}{3}$ = $22.88$ 
\end{center}
Hence, the result obtained from sequence1, which contains 60-dimensional numeric feature vectors is as follows: 
$[2, 28, 169.0, 2, 76, 1.0, 1, 19, 0.0, 2, 47, 6.25, 1, 34, 0.0, 3, 28,\\ 22.8, 2, 18, 36.0, 2, 28, 16.0, 4, 64, 127.5, 3, 100, 64.8, 0, 0, 0, 2,\\ 45, 110.25, 4, 75, 69.6, 1, 36, 0.0, 1, 32, 0.0, 1, 30, 0.0, 4, 31,\\ 35.1, 5, 106, 108.2, 1, 23, 0.0, 0, 0, 0]$

The results of all the other four sequences are shown in Fig. \ref{sequence-result}.

\subsection{Scalable 6-dimensional Feature Extraction Approach}
\label{SCPSF}
\begin{figure}[ht]
  \includegraphics[width=0.5\textwidth]{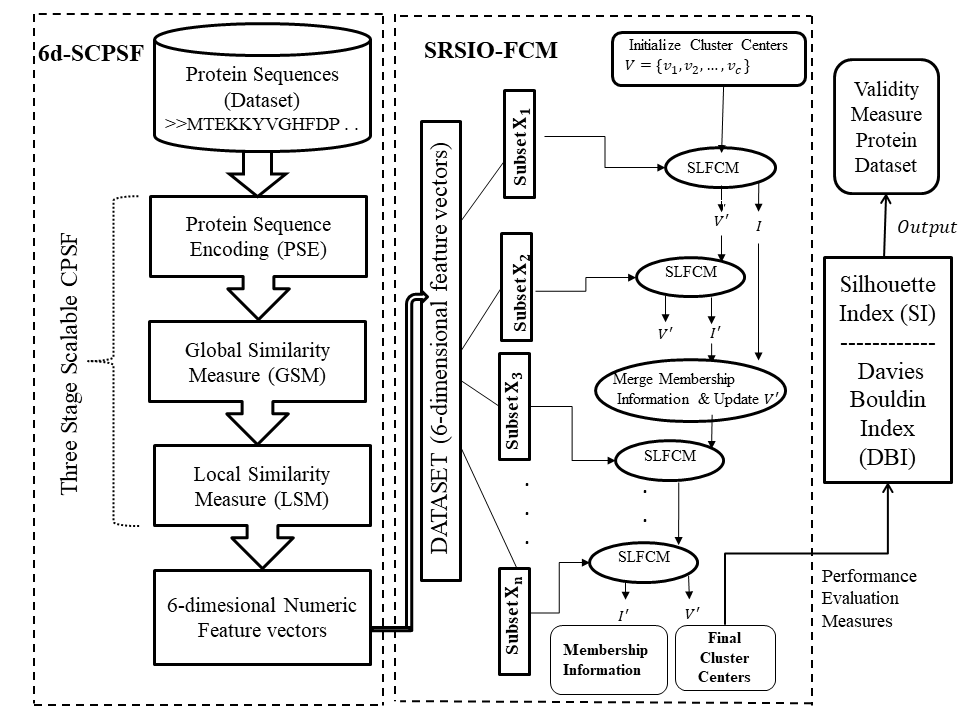}
\caption{6d-SCPSF architecture embedded using SRSIO-FCM and SLFCM with performance measure evaluation.}
\label{6d-architecture}      
\end{figure}

The proposed scalable feature extraction approach 6d-SCPSF  is being implemented using Apache Spark framework to represent all the protein sequences in 6-dimensional numeric feature vectors.   The architecture of the proposed 6d-SCPSF extraction approach is shown in Figure \ref{6d-architecture}. 
Algorithm \ref{alg:6d-scpsf} summarizes the steps of 6d-SCPSF approach.
The input given is a raw protein dataset containing 20 amino acids $\sum$=\{A, C, D, E, F, G, H, I, K, L, M, N, P, Q, R, S, T, V, W, Y\}. The output is a feature vector of the given input dataset, which is a file containing 6-dimensional numeric feature vectors. 
\begin{algorithm}[t]
	\caption{6d-SCPSF algorithm}
	\label{alg:6d-scpsf}
\begin{algorithmic}[1]
\REQUIRE Raw protein data : $raw\_Protein.txt$
	\ENSURE Preprocessed protein data : $ pre\_Protein.txt$
 \STATE \textbf{Call} Scalable Protein Sequence Encoding (SPSE) algorithm.
 \STATE \textbf{Call} Global Similarity Matrix (GSM) algorithm. 
 \STATE \textbf{Call} Scalable Local Similarity Measures (SLSM) algorithm. 
 \STATE \textbf{Save} as text file $(Feature\_Vectors.txt)$ 
\end{algorithmic}
\end{algorithm}
The 6d-SCPSF extraction approach is explained in Algorithm \ref{alg:6d-scpsf}. Line 1 of Algorithm \ref{alg:6d-scpsf}, calls the SPSE algorithm, which distributes $raw\_Protein.txt$ protein dataset on Apache Spark clusters.  
In Line 2, the GSM algorithm is called, which calculates the probability matrix of the sequences on a master machine without distributing the dataset in spark clusters. 
The output obtained from the Scalable Protein Sequence Encoding (SPSE)  algorithm is used as an input to the GSM algorithm.  In Line 3, the Scalable Local Similarity Measures (SLSM) algorithm is called, which again distributes the output obtained from the GSM algorithm using the Apache Spark cluster.
Finally, the 6-dimensional numerical feature vectors are saved in a file using Line 4. 
The subsequent section explains the SPSE, GSM, and SLSM algorithms.
\subsubsection{SPSE Algorithm}
The PSE algorithm encodes each amino acid into particular encoding groups, as discussed in Section \ref{stage1}. This section presents the scalable version of PSE (SPSE). Algorithm~\ref{alg:spse} shows the steps of the SPSE approach.
In Line 1, data is read into a resilient distributed dataset (RDD) from Hadoop.  In Line 2, the file ($raw\_Protein.txt$) is passed to the Encode function and store as the return values in other Spark RDD. Then the map function distributes Encode() method to every worker node for parallel execution of the task in Line 3. In Line 4, we save the data of the encoded RDD to the text file.   From Line 5-12, the work of Encode() function is given. The amino acids are mapped with the exchange groups from Line 7-12. 
The DataFrame is used to store the computation of each step.  
\begin{algorithm}
	\caption{Scalable Protein Sequence Encoding (SPSE)}
	\label{alg:spse}
\begin{algorithmic}[1]
\REQUIRE Raw protein sequence: $raw\_Protein.txt$
  \ENSURE Encoded protein sequence: $enc\_Protein.txt$
 \STATE \textbf{Read} the protein sequences from the file and parallelize them with the help of Spark RDD.
  \STATE \textbf{Pass} the file ($raw\_Protein.txt$) to \textbf{Encode} function and store the return values in other Spark RDD. 
  \STATE Map Encode function.
   \STATE \textbf{Save} data of encoded RDD to text file. 
  \STATE  \textit{Function} Encode(z)$\{$ 
 \STATE   \textbf{Store} the data in the DataFrame and split each letter in the sequence to different columns. 
 \STATE  \textbf{Replace} the $'H','R','K'$ with $e_1$. 
  \STATE \textbf{Replace} the $'D','E','N','Q'$ with $e_2$. 
 \STATE  \textbf{Replace} the $'C'$ with $e_3$. 
 \STATE  \textbf{Replace} the $'S','T','P','A','G'$  with $e_4$. 
  \STATE \textbf{Replace} the $'M','I','L','V'$  with $e_5$. 
 \STATE  \textbf{Replace} the $'F','Y','W'$  with $e_6$. 
 \STATE  \textbf{Returns} the encoded protein sequence: $ enc\_Protein.txt$ $\}$ 
  \end{algorithmic}
\end{algorithm}
In Spark, a DataFrame is a distributed collection of data organized into named columns.  In Line 13, the output of the SPSE algorithm is saved in the $enc\_Protein.txt$ file. After that, the $enc\_Protein.txt$ file is taken as input to the GSM algorithm, which is presented next.

\subsubsection{GSM Algorithm}
The GSM is used to calculate the instance probability of all exchange groups at each position relative to the total number of protein sequences. Section \ref{stage2} discusses the steps used to find global similarity measures of the protein-encoding sequence. Algorithm~\ref{alg:gsm} discusses the steps of the GSM approach. The input of this algorithm is $enc\_Protein.txt$, which is obtained from Algorithm~\ref{alg:spse}. 
Line 1 reads the data from a file and saves it to a DataFrame. Line 2 finds the number of columns in the DataFrame, and Line 3 locates the number of rows in the DataFrame. An empty DataFrame is created for the estimation of the probability DataFrame of sequences in Line 4.
\begin{algorithm}
	\caption{Global Similarity Measures (GSM)}
	\label{alg:gsm}
\begin{algorithmic}[1]
\REQUIRE Encoded protein sequences: $ enc\_Protein.txt$
  \ENSURE Probability DataFrame of sequences: $prob\_Protein.txt$ 
\STATE \textbf{Read} the encoded sequences data from the file and store it in the DataFrame.
\STATE    \textbf{Find} the number of columns of the DataFrame(col). 
 \STATE    \textbf{Find} the number of rows of the DataFrame(row).
\STATE    \textbf{Create} an empty probability DataFrame with index names $e_1,e_2,e_3,e_4,e_5,e_6$.
\FOR{ column in range 0 to col}
  \STATE \textbf{Create} an empty list.
\FOR{exchange group }
 \STATE  \textbf{Get} the occurrence of exchange group in that column.
 \STATE  \textbf{Get} the probability of the exchange group using Eq. (\ref{eq:E9}).
  \STATE  \textbf{Add} probability to the list.
\ENDFOR
\STATE   \textbf{Add} list as a new column in the probability DataFrame.
\ENDFOR
  \STATE   \textbf{Save} probability DataFrame as text file $(prob\_Protein.txt)$. 
\end{algorithmic}
\end{algorithm}
Then, calculate the occurrence of the exchange group for each column and add this to the DataFrame in Line 8.  In Line 9, the value of the probability of the exchange group is computed using Eq. (\ref{eq:E9}). Line 10 adds the probability to the list. In Line 12, a list is added as a new column in the probability DataFrame. Line 14 saves the result to $prob\_Protein.txt$ file used as input to the SLSM algorithm.    After that, $enc\_Protein.txt$  and $prob\_Protein.txt$ files are taken as input to the next stage in the SLSM algorithm, which is presented next.

\subsubsection{SLSM Algorithm}

This section presents the scalable version of LSM (SLSM). The SLSM determines the location-specific weight of each exchange group within the sequence and produces the weight factors. These weight factors ultimately represent the numeric feature vectors for each protein sequence. Section \ref{stage3} discusses the steps used to find local similarity measures of the protein-encoding sequence. 
Algorithm~\ref{alg:slsm} discusses the steps of the SLSM approach.
\begin{algorithm}
 	\caption{Scalable Local Similarity Measures (SLSM)}
	\label{alg:slsm}
\begin{algorithmic}[1]
\REQUIRE Probability matrix of sequences: $prob\_Protein.txt$, encoded protein sequences: $enc\_Protein.txt$
  \ENSURE Feature Vectors: $Feature\_Vectors.txt$
 
\STATE \textbf{Read} the sequences from the file $(enc\_Protein.txt)$ and parallelize using Spark RDD.
  \STATE  \textbf{Read} the data from the $prob\_Protein.txt$ and store it in DataFrame (probability).
\STATE    Map FeatureVector function. 
\STATE    \textbf{Save} data of FeatureVector RDD to text file.
\STATE    \textit{Function} \textbf{FeatureVector}(probability, data) {
 \STATE   \textbf{Create} an empty featureVector DataFrame with column names as {$e_1,e_2,e_3,e_4,e_5,e_6$}.
 \STATE   \textbf{Create} DataFrame using data.
  \STATE  \textbf{Find} the number of columns of the DataFrame(cols).
\STATE    \textbf{Find} the number of rows of the DataFrame(rows).
\FOR{ row in range of 0 to rows}
\STATE    \textbf{Create} a dictionary as featureVector = { $'e_1':0,'e_2':0,'e_3':0,'e_4':0,'e_5':0,'e_6':0$}. 
\FOR{ col in range of 0 to cols}
\STATE    \textbf{Get} the exchange group from the position of rows and cols in DataFrame.
\STATE    \textbf{Modify} the key value of the exchange group from featureVector by adding previous value with the value of probability at the position of exchange group and cols from probability DataFrame. 
\ENDFOR
  \STATE   \textbf{Add} the featureVector to the featureVector DataFrame.
\ENDFOR }
   \STATE \textbf{Save} the featureVector DataFrame to text file $(Feature\_Vectors.txt)$.
   \end{algorithmic}
\end{algorithm}

This algorithm takes input from two files, i.e., $enc\_Protein.txt$ and $prob\_Protein.txt$, which are obtained from Algorithm~\ref{alg:spse} and Algorithm~\ref{alg:gsm}, respectively.  The data ($enc\_Protein.txt$) is read into an RDD from Hadoop in Line 1. Then, the data from the file ($prob\_Protein.txt$) is accessed and stored in DataFrame (probability) in Line 2. The map function distributes FeatureVector() method to every worker node for parallel execution of the task in Line 3.  In Line 4, we save the data obtained from the featureVector RDD to a text file. From Line 5-17, the working of the FeatureVector() function is given, where the DataFrame is used to store the computation of each step.  An empty DataFrame is created with the given column names in Line 6. Then, a DataFrame is created using the data in Line 7. Line 8 finds the number of cols, and Line 9 locates the number of rows of the DataFrame. Line 11 creates a dictionary of feature vectors. The exchange group of rows and columns from DataFrame is acquired in line 13.  In Line 14, we modify the key value of the exchange group from featureVector. This is done by adding previous value with the value of probability at the position of the exchange group and columns from the probability DataFrame. Line 10-17  is computed using Eq. (\ref{eq:E10}).  Finally, the results are saved in the $Feature\_Vectors.txt$ file on Line 18.

The proposed scalable feature extraction algorithm (6d-SCPSF) has the significant characteristic that it takes raw protein sequences as input and produces 6-dimensional numeric feature vectors as output using the Apache Spark framework. The output produced from the proposed feature extraction approaches in terms of 60-dimensional and 6-dimensional feature vectors is passed as an input to the SRSIO-FCM and SLFCM clustering algorithms to produce output in terms of clusters consisting of protein sequences. The output produced by the clustering algorithms is reported in terms of clusters.

\section{Experimental results} \label{sec4}
In the experiments, we analyzed the performance of the proposed 60d-SPF and 6d-SCPSF extraction methods, which are applied to both the SRSIO-FCM and SLFCM algorithms. Their performance is measured in terms of the Silhouette index (SI) and Davies Bouldin index (DBI).
Furthermore, feature vectors using Bandyopadhyay \cite{bandyopadhyay2005efficient}  and Mansoori \cite{mansoori2009protein} methods have been computed. Then, the clustering of these feature vectors is carried out using the SRSIO-FCM and SLFCM algorithms. The performance of the proposed 60d-SPF and 6d-SCPSF is compared with that of Bandyopadhyay and Mansoori in terms of SI and DBI, respectively.

\subsection{Experimental environment}
We have implemented the proposed algorithms on the Apache Spark cluster. The Spark cluster consists of one master and five worker nodes. The master node is equipped with 32 GB of RAM, 8 cores, and 3 TB of storage. Each worker node is equipped with 16 GB of RAM, 8 cores, and 1 TB of storage. 
\subsection{Dataset description}
We have investigated the performance of the proposed 60d-SPF and 6d-SCPSF approaches compared with those of Bandyopadhyay and Mansoori on SRSIO-FCM and SLFCM algorithms on the following protein datasets. A detailed description of the protein datasets used for the experimental analysis is present in Table  \ref{parameters}. 

\begin{table}[t]
\caption{Description of soybean protein datasets.}
\resizebox{\columnwidth}{!}{
\label{parameters}
\begin{tabular}{lllll}
\noalign{\smallskip}\hline\noalign{\smallskip}\noalign{\smallskip}
\multicolumn{1}{l}{\textbf{Parameters} } & \multicolumn{4}{l}{\textbf{Datasets}} 
 \\\noalign{\smallskip}
\cline{2-5} \noalign{\smallskip}
                            \multicolumn{1}{l}{}                                                             &               \multicolumn{1}{l}{Lee}     & \multicolumn{1}{l}{Williams82} &
                           \multicolumn{1}{l}{PI483463} &
                             \multicolumn{1}{l}{W05}
                           \\
 
\noalign{\smallskip}\hline\noalign{\smallskip}
\textbf{sequences}      & 71358    & 73320      & 62102  & 89477\\

\textbf{size}      & 44 MB      & 53.4 MB      & 38.9 MB  & 50.7 MB\\
\noalign{\smallskip}\hline\noalign{\smallskip}\noalign{\smallskip}
\end{tabular}}
\end{table}

\begin{table}[]
\caption{Comparison of the number of features extracted with different feature extraction approaches.}
\renewcommand{\arraystretch}{1.3}
\label{comparison}
\begin{tabular}{lll}
\hline
Authors         & Feature Extraction Method      & Features \\ \hline
Bandyopadhyay   & 1-gram feature encoding method & 20                 \\
Mansoori  & Exchange group encoding method & 6                  \\
Proposed 6d-SCPSF        & Global and local similarity    & 6      \\       Proposed 60d-SPF  &   Pseudo amino acid composition & 60 \\
\hline
\end{tabular}
\end{table}

\subsubsection{Lee}
\noindent
The Lee strain, which crosses between the Chinese lines CNS and $S$-$100$, is widely used as a parent in many breeding projects in the southern United States and Brazil. Diversity is characterized by resistance to bacteria from Phytophthora rot, Peanut Mottle Virus, and bacterial pustule \cite{Lee}.
\begin{table*}[t]
\renewcommand{\arraystretch}{1.3}

\caption{SI values of LEE protein dataset applied on SRSIO-FCM and SLFCM algorithms.}
\label{si-Lee}
\centering
\begin{tabular}{lllllllll}

\hline
\multirow{3}{*}{clusters} &
  \multicolumn{8}{l}{Algorithms} \\ \cline{2-9} 
 &
  \multicolumn{4}{l}{SRSIO-FCM} &
  \multicolumn{4}{l}{SLFCM} \\ \cline{2-9} 
 &
  \multicolumn{1}{l}{60d-SPF} &
  \multicolumn{1}{l}{6d-SCPSF} &
  \multicolumn{1}{l}{Bandyopadhyay} &
  \multicolumn{1}{l}{Mansoori} &
  \multicolumn{1}{l}{60d-SPF} &
  \multicolumn{1}{l}{6d-SCPSF} &
  \multicolumn{1}{l}{Bandyopadhyay} &
  Mansoori \\ \hline
5                   & 0.6678 & 0.4501 & 0.1321  & 0.1327 & 0.7232 & 0.4506 & 0.1683  & 0.1469 \\
10                  & 0.4826 & 0.3361 & 0.0461  & 0.0772 & 0.4802 & 0.3361 & 0.0572  & 0.0501 \\
15                  & 0.3922 & 0.2721 & 0.0129  & 0.0703 & 0.4039 & 0.2706 & 0.0184  & 0.0556 \\
20                  & 0.3522 & 0.2511 & -3.6947 & 0.0447 & 0.3505 & 0.2513 & 0.0007  & 0.0464 \\
25                  & 0.3074 & 0.2303 & -0.0067 & 0.0369 & 0.3118 & 0.2329 & -0.0006 & 0.0516 \\
30                  & 0.2131 & 0.199  & -0.0143 & 0.0357 & 0.2769 & 0.2164 & -0.0088 & 0.0482
   \\ \hline
\end{tabular}
\end{table*}

\begin{table*}[t]
\renewcommand{\arraystretch}{1.3}

\caption{SI values of Williams82 protein dataset applied on SRSIO-FCM and SLFCM algorithms.}
\label{si-Wm82}
\centering
\begin{tabular}{lllllllll}

\hline
\multirow{3}{*}{clusters} &
  \multicolumn{8}{l}{Algorithms} \\ \cline{2-9} 
 &
  \multicolumn{4}{l}{SRSIO-FCM} &
  \multicolumn{4}{l}{SLFCM} \\ \cline{2-9} 
 &
  \multicolumn{1}{l}{60d-SPF} &
  \multicolumn{1}{l}{6d-SCPSF} &
  \multicolumn{1}{l}{Bandyopadhyay} &
  \multicolumn{1}{l}{Mansoori} &
  \multicolumn{1}{l}{60d-SPF} &
  \multicolumn{1}{l}{6d-SCPSF} &
  \multicolumn{1}{l}{Bandyopadhyay} &
  Mansoori \\ \hline 
5                   & 0.6672 & 0.4607 & 0.0348  & 0.1234 & 0.8021 & 0.4999 & 0.0279  & 0.1278 \\
10                  & 0.5055 & 0.3391 & -0.026  & 0.089  & 0.5758 & 0.3728 & -0.0272 & 0.0689 \\
15                  & 0.4174 & 0.2706 & -0.047  & 0.0612 & 0.4676 & 0.293  & -0.0277 & 0.0467 \\
20                  & 0.3918 & 0.2538 & -0.0835 & 0.0475 & 0.4122 & 0.2569 & -0.0681 & 0.0429 \\
25                  & 0.331  & 0.2252 & -0.1142 & 0.0398 & 0.3916 & 0.2399 & -0.066  & 0.0346 \\
30                  & 0.2868 & 0.2201 & -0.1153 & 0.0341 & 0.3548 & 0.2267 & -0.0714 & 0.0259
   \\ \hline
\end{tabular}
\end{table*}

\begin{table*}[t]
\renewcommand{\arraystretch}{1.3}

\caption{SI values of PI48346 protein dataset applied on SRSIO-FCM and SLFCM algorithms.}
\label{si-PI}
\centering
\begin{tabular}{lllllllll}

\hline
\multirow{3}{*}{clusters} &
  \multicolumn{8}{l}{Algorithms} \\ \cline{2-9} 
 &
  \multicolumn{4}{l}{SRSIO-FCM} &
  \multicolumn{4}{l}{SLFCM} \\ \cline{2-9} 
 &
  \multicolumn{1}{l}{60d-SPF} &
  \multicolumn{1}{l}{6d-SCPSF} &
  \multicolumn{1}{l}{Bandyopadhyay} &
  \multicolumn{1}{l}{Mansoori} &
  \multicolumn{1}{l}{60d-SPF} &
  \multicolumn{1}{l}{6d-SCPSF} &
  \multicolumn{1}{l}{Bandyopadhyay} &
  Mansoori \\
  \hline
5                   & 0.6329 & 0.4492 & 0.0498  & 0.1154 & 0.6283 & 0.4493 & 0.0594  & 0.1818 \\
10                  & 0.478  & 0.3358 & 0.0203  & 0.0665 & 0.4766 & 0.3358 & 0.0166  & 0.0863 \\
15                  & 0.3894 & 0.2533 & -0.0349 & 0.0593 & 0.3919 & 0.2557 & -0.0036 & 0.0605 \\
20                  & 0.3471 & 0.24   & -0.0759 & 0.0533 & 0.3393 & 0.2477 & -0.075  & 0.0445 \\
25                  & 0.3063 & 0.2244 & -0.1174 & 0.039  & 0.3065 & 0.2275 & -0.0306 & 0.0509 \\
30                  & 0.2699 & 0.2013 & -0.1556 & 0.0366 & 2.0357 & 0.2219 & -0.1008 & 0.0561
   \\ \hline
\end{tabular}
\end{table*}


\begin{table*}[]
\renewcommand{\arraystretch}{1.3}

\caption{SI values of W05 protein dataset applied on SRSIO-FCM and SLFCM algorithms.}
\label{si-W05}
\centering
\begin{tabular}{lllllllll}

\hline
\multirow{3}{*}{clusters} &
  \multicolumn{8}{l}{Algorithms} \\ \cline{2-9} 
 &
  \multicolumn{4}{l}{SRSIO-FCM} &
  \multicolumn{4}{l}{SLFCM} \\ \cline{2-9} 
 &
  \multicolumn{1}{l}{60d-SPF} &
  \multicolumn{1}{l}{6d-SCPSF} &
  \multicolumn{1}{l}{Bandyopadhyay} &
  \multicolumn{1}{l}{Mansoori} &
  \multicolumn{1}{l}{60d-SPF} &
  \multicolumn{1}{l}{6d-SCPSF} &
  \multicolumn{1}{l}{Bandyopadhyay} &
  Mansoori \\
  \hline 
5                   & 0.6505  & 0.4526 & 0.1706 & 0.0567 & 0.7219  & 0.4676 & 0.1668  & 0.1658 \\
10                  & 0.4895  & 0.3341 & 0.066  & 0.0384 & 0.51031 & 0.3409 & 0.0825  & 0.1707 \\
15                  & 0.4063  & 0.2732 & 0.025  & 0.0383 & 0.4519  & 0.3136 & 0.0371  & 0.1298 \\
20                  & 0.36365 & 0.2494 & 0.0078 & 0.0286 & 0.3898  & 0.2618 & 0.0175  & 0.0947 \\
25                  & 0.3087  & 0.2358 & 0.0103 & 0.0334 & 0.3447  & 0.2467 & 0.0029  & 0.0692 \\
30                  & 0.2719  & 0.217  & 0.0004 & 0.0309 & 0.3023  & 0.2325 & -0.0081 & 0.0486
   \\ \hline
\end{tabular}
\end{table*}

\begin{table*}[t]
\renewcommand{\arraystretch}{1.3}

\caption{DBI values of LEE protein dataset applied on SRSIO-FCM and SLFCM algorithms.}
\label{dbi-Lee}
\centering
\begin{tabular}{lllllllll}

\hline
\multirow{3}{*}{clusters} &
  \multicolumn{8}{l}{Algorithms} \\ \cline{2-9} 
 &
  \multicolumn{4}{l}{SRSIO-FCM} &
  \multicolumn{4}{l}{SLFCM} \\ \cline{2-9} 
 &
  \multicolumn{1}{l}{60d-SPF} &
  \multicolumn{1}{l}{6d-SCPSF} &
  \multicolumn{1}{l}{Bandyopadhyay} &
  \multicolumn{1}{l}{Mansoori} &
  \multicolumn{1}{l}{60d-SPF} &
  \multicolumn{1}{l}{6d-SCPSF} &
  \multicolumn{1}{l}{Bandyopadhyay} &
  Mansoori \\
  \hline
5                   & 0.6654 & 0.7227 & 2.125  & 1.443  & 0.536  & 0.7223 & 1.8179 & 1.7073 \\
10                  & 0.8207 & 0.9299 & 3.5887 & 1.7254 & 0.8214 & 0.9332 & 4.1065 & 1.8161 \\
15                  & 1.1391 & 1.1504 & 5.445  & 1.7718 & 1.1412 & 1.1535 & 5.8585 & 1.8317 \\
20                  & 1.382  & 1.1829 & 6.5559 & 1.8216 & 1.4467 & 1.1468 & 6.4553 & 1.9362 \\
25                  & 1.7125 & 1.1861 & 7.6284 & 1.8225 & 0.6562 & 1.1641 & 7.2828 & 1.8729 \\
30                  & 1.7456 & 1.2772 & 8.7204 & 1.8378 & 1.8213 & 1.1876 & 7.9131 & 1.8648 
   \\ \hline
\end{tabular}
\end{table*}

\begin{table*}[t]
\renewcommand{\arraystretch}{1.3}

\caption{DBI values of Williams82 protein dataset applied on SRSIO-FCM and SLFCM algorithms.}
\label{dbi-Wm82}
\centering
\begin{tabular}{lllllllll}

\hline
\multirow{3}{*}{clusters} &
  \multicolumn{8}{l}{Algorithms} \\ \cline{2-9} 
 &
  \multicolumn{4}{l}{SRSIO-FCM} &
  \multicolumn{4}{l}{SLFCM} \\ \cline{2-9} 
 &
  \multicolumn{1}{l}{60d-SPF} &
  \multicolumn{1}{l}{6d-SCPSF} &
  \multicolumn{1}{l}{Bandyopadhyay} &
  \multicolumn{1}{l}{Mansoori} &
  \multicolumn{1}{l}{60d-SPF} &
  \multicolumn{1}{l}{6d-SCPSF} &
  \multicolumn{1}{l}{Bandyopadhyay} &
  Mansoori \\
  \hline
5                   & 0.603 & 0.7409 & 3.2884 & 1.6421 & 0.6196 & 0.691  & 2.9909 & 1.7052 \\
10                  & 0.7689 & 1.0339 & 5.2884 & 1.7618 & 0.6086 & 0.842  & 7.5041 & 1.9261 \\
15                  & 1.0555 & 1.2841 & 6.9123 & 1.9205 & 0.8371 & 1.1241 & 6.7961 & 1.9288 \\
20                  & 1.1908 & 1.2863 & 8.4169 & 1.9681 & 1.0746 & 1.1763 & 7.7635 & 1.8761 \\
25                  & 1.141  & 1.2663 & 8.8731 & 1.8905 & 0.9778 & 1.2336 & 8.1169 & 1.8983 \\
30                  & 1.4429 & 1.3602 & 8.0659 & 1.8505 & 1.0743 & 1.2645 & 7.6    & 1.9418
   \\ \hline
\end{tabular}
\end{table*}
\begin{table*}[t]
\renewcommand{\arraystretch}{1.3}

\caption{DBI values of PI48346 protein dataset applied on SRSIO-FCM and SLFCM algorithms.}
\label{dbi-PI}
\centering
\begin{tabular}{lllllllll}

\hline
\multirow{3}{*}{clusters} &
  \multicolumn{8}{l}{Algorithms} \\ \cline{2-9} 
 &
  \multicolumn{4}{l}{SRSIO-FCM} &
  \multicolumn{4}{l}{SLFCM} \\ \cline{2-9} 
 &
  \multicolumn{1}{l}{60d-SPF} &
  \multicolumn{1}{l}{6d-SCPSF} &
  \multicolumn{1}{l}{Bandyopadhyay} &
  \multicolumn{1}{l}{Mansoori} &
  \multicolumn{1}{l}{60d-SPF} &
  \multicolumn{1}{l}{6d-SCPSF} &
  \multicolumn{1}{l}{Bandyopadhyay} &
  Mansoori \\
  \hline
5                   & 0.6434 & 0.7243 & 2.6845 & 1.7117 & 0.6439 & 0.7249 & 2.8005 & 1.494  \\
10                  & 0.8636 & 0.9302 & 5.2577 & 1.7012 & 0.864  & 0.9311 & 5.3841 & 1.7239 \\
15                  & 1.117  & 1.2587 & 5.8198 & 1.8244 & 1.1296 & 1.2434 & 7.4839 & 1.7644 \\
20                  & 1.1223 & 1.2611 & 7.2405 & 1.8055 & 1.5365 & 1.1831 & 5.918  & 1.7856 \\
25                  & 1.176  & 1.244  & 6.934  & 1.8702 & 1.8109 & 1.1824 & 7.2622 & 1.7718 \\
30                  & 1.7778 & 1.3    & 6.6071 & 1.8825 & 0.2883 & 1.2268 & 8.4102 & 1.8205
 
   \\ \hline
\end{tabular}
\end{table*}
\begin{table*}[t]
\renewcommand{\arraystretch}{1.3}

\caption{DBI values of W05 protein dataset applied on SRSIO-FCM and SLFCM algorithms.}
\label{dbi-W05}
\centering
\begin{tabular}{lllllllll}

\hline
\multirow{3}{*}{clusters} &
  \multicolumn{8}{l}{Algorithms} \\ \cline{2-9} 
 &
  \multicolumn{4}{l}{SRSIO-FCM} &
  \multicolumn{4}{l}{SLFCM} \\ \cline{2-9} 
 &
  \multicolumn{1}{l}{60d-SPF} &
  \multicolumn{1}{l}{6d-SCPSF} &
  \multicolumn{1}{l}{Bandyopadhyay} &
  \multicolumn{1}{l}{Mansoori} &
  \multicolumn{1}{l}{60d-SPF} &
  \multicolumn{1}{l}{6d-SCPSF} &
  \multicolumn{1}{l}{Bandyopadhyay} &
  Mansoori \\ 
  \hline
5                   & 0.6563 & 0.7259 & 1.867  & 1.8148 & 0.5976 & 0.7317 & 1.8314 & 1.6565 \\
10                  & 0.8644 & 1.0193 & 4.1775 & 2.0667 & 0.7805 & 0.9628 & 3.5178 & 1.4468 \\
15                  & 1.0727 & 1.1937 & 6.2512 & 1.9356 & 0.9015 & 1.0197 & 4.8952 & 1.5505 \\
20                  & 1.3559 & 1.2546 & 7.3401 & 1.9127 & 1.0317 & 1.1793 & 4.4736 & 1.9088 \\
25                  & 1.7617 & 1.1998 & 8.0035 & 2.0381 & 1.4231 & 1.1975 & 6.3209 & 1.8746 \\
30                  & 1.7758 & 1.2694 & 9.0045 & 1.9626 & 1.7899 & 1.1922 & 7.5768 & 1.8836
   \\ \hline
\end{tabular}
\end{table*}

\subsubsection{Williams82}
\noindent
Williams 82, a soybean cultivar used to construct the reference genome sequence, was created by reversing the Phytophthora root rot resistance locus from the donor parent Kingwa to the recurrent Williams parent \cite{2017soybean}.
\subsubsection{PI483463}
\noindent
\textit{Glycine Soja} is the closest wild soybean of Glycine max. Species remain interfertile, and specimens of G. soja are used in breeding projects to introduce traits such as resistance to certain diseases or environmental stress. Glycine swelling accession PI483463 is known to be abnormally tolerable.  The genome of this entry is sequential and is partly based on this salt tolerance \cite{PI}. 
\subsubsection{W05}
\noindent
\textit{Glycine Soja} accession W05 is a salt-tolerant wild soybean whose genome is designed to serve as a reference genome assembly. W05 affiliation has been used for genetic studies of various traits including uncertainty, seed size, number of pods per plant, and seed color \cite{W05}. 

All the protein datasets discussed in this section are available  at the following URL: https://soybase.org/dlpages/.
\subsection{Performance evaluation}
\subsubsection{Silhouette index (SI)}
SI is a metric that compares how similar a data sample is to its cluster to other clusters. The
Silhouette value is limited to a number between -1 and 1. A negative number implies poor clustering quality, whereas a positive value suggests excellent clustering quality \cite{silhouette}.
Thus SI is characterized as:
\begin{equation}
S(i) = \frac{a_2(i) - a_1(i)} { max [a_1(i), a_2(i)]}
\end{equation}
Where $a_1(i)$ is the average distance between $i^{th}$ sample from all other data samples within the same cluster, $a_2(i)$ is the lowest average distance of $i^{th}$ sample to all the data samples in any other cluster, of which $i$ is not a member.
\subsubsection{Davies-Bouldin index (DBI)}
DBI   divides a single
record into two measures, one for the dispersion of individual
clusters and the other for the partitioning of distinct clusters \cite{dbs}.
The DBI is not constrained inside a particular range, thus a lower DBI implies higher clustering quality. 
Thus, DBI is characterized as:
\begin{equation}
   DBI = {1 \over c}{\sum}_{i=1}^{c} {\max}_{j \neq i}\left[{{\rm
diam}(C_{i})+{\rm diam}(C_{j}) \over {\rm d}(C_{i},C_{j})}\right]
\end{equation}
Where $d(C_i,C_j)$ correlates to the distance between the center of clusters $C_i$ and $C_j$, diam $(C_i)$ is the maximum distance between all the data samples of cluster $C_i$ and $c$ is the number of clusters.

\subsection{Results and discussion} 
In this section, we discuss the effectiveness of the proposed 60d-SPF compared with the proposed 6d-SCPSF extraction and existing methods like Bandyopadhyay and Mansoori when applied to SRSIO-FCM and SLFCM algorithms on protein data, and the results are investigated in terms of SI and DBI. The SRSIO-FCM algorithm partitions the protein dataset into three subsets, where the subsets are the subsets of the entire data. However, SLFCM works over the whole data. The clustering is performed on clusters 5, 10, 15, 20, 25, and 30, respectively. The performance of the proposed feature extraction approach is evaluated in comparison with other feature extraction approaches discussed in the literature. The feature encoding techniques used here for comparison are reported along with the number of features extracted by each approach in Table \ref{comparison}. The comparison of proposed 60d-SPF and 6d-SCPSF, existing feature extraction methods like Bandyopadhyay and Mansoori applied to SRSIO-FCM, and SLFCM algorithms on all four protein datasets is shown in tables reported subsequently.

 Tables \ref{si-Lee}-\ref{si-W05}  highlight SI values for the Lee, Williams82, PI483463, and W05 soybean protein datasets, which are computed after supplying the features extracted from the proposed 60d-SPF and 6d-SCPSF extraction and existing methods like Bandyopadhyay and Mansoori into the SRSIO-FCM and SLFCM clustering algorithms. SI demonstrates the quality of clustering. The higher SI value indicates good clustering results.  Table \ref{si-Lee} shows the SI values for the Lee dataset. The SRSIO-FCM and SLFCM algorithms have obtained a lower value for the 6d-SCPSF extraction method than the 60d-SPF extraction method. On the other hand, the existing methods (Bandyopadhyay and Mansoori) have significantly lower values than those proposed. Observing the SI values, the SRSIO-FCM and SLFCM algorithms have obtained negative values for most of the clusters for Bandyopadhyay and extremely low for Mansoori. Moreover, for the 60d-SPF approach, the SRSIO-FCM and SLFCM achieved the highest value of SI for cluster 5. 
 
Table \ref{si-Wm82} shows the SI values for the Williams82. The SRSIO-FCM and SLFCM algorithms have obtained a lower value for the 6d-SCPSF extraction method than the 60d-SPF extraction method. On the other hand, the existing methods (Bandyopadhyay and Mansoori) have significantly lower values than those proposed.  Observing the SI values, the SRSIO-FCM and SLFCM algorithms have obtained negative values for most of the clusters for Bandyopadhyay and extremely low for Mansoori. Additionally, for the 60d-SPF approach, the SRSIO-FCM and SLFCM achieved the highest value of SI for cluster 5. 

Table \ref{si-PI} shows the SI values for the PI483463 dataset. The SRSIO-FCM and SLFCM algorithms have obtained a higher value for the 60d-SPF extraction method than the 6d-SCPSF extraction method. On the other hand, the existing methods (Bandyopadhyay and Mansoori) have significantly lower values than those proposed.  Observing the SI values, the SRSIO-FCM and SLFCM algorithms have obtained negative values for most of the clusters for Bandyopadhyay and extremely low for Mansoori. Additionally, for the 60d-SPF approach, the SRSIO-FCM and SLFCM achieved the highest value of SI for cluster 5. 

Table \ref{si-W05} shows the SI values for the W05 dataset. The SRSIO-FCM and SLFCM algorithms have obtained a higher value for the 60d-SPF extraction method than the 6d-SCPSF extraction method. On the other hand, the existing methods (Bandyopadhyay and Mansoori) have significantly lower values than those proposed.  Observing the SI values, the SRSIO-FCM and SLFCM algorithms have obtained extremely low  values for most of the clusters for Bandyopadhyay and Mansoori. Additionally, for the 60d-SPF approach, the SRSIO-FCM and SLFCM achieved the highest value of SI for cluster 5.  

Table \ref{dbi-Lee}-\ref{dbi-W05} highlights  DBI values for the Lee, Williams82, PI483463, and W05 soybean protein datasets. Conversely to the SI,  the DBI is not bounded within a given range. As a general rule, the lower the DBI value, the better the clustering result.  Table \ref{dbi-Lee} shows the DBI values for the Lee dataset. The value achieved by SRSIO-FCM and SLFCM is much better for the 60d-SPF than 6d-SCPSF. As we can see, for 60d-SPF, the DBI values are lower than 6d-SCPSF when clustered using SRSIO-FCM and SLFCM on almost all the clusters. On the other hand, the existing methods (Bandyopadhyay and Mansoori) have extremely higher values than those proposed.  Moreover, SRSIO-FCM and SLFCM attained a very low value in cluster 5.  

 Table \ref{dbi-Wm82} shows the DBI values for the Williams82 dataset. The value achieved by SRSIO-FCM and SLFCM is much better for the 60d-SPF than 6d-SCPSF. As we can see, for the 6d-SCPSF approach, the DBI values are higher than the 60d-SPF approach when clustered using SRSIO-FCM and SLFCM on all the clusters except cluster 30 for SRSIO-FCM. On the other hand, the existing methods (Bandyopadhyay and Mansoori) have extremely higher values than those proposed.  Moreover, SRSIO-FCM attained a low value on cluster 5 for 60d-SPF. 

 Table \ref{dbi-PI} shows the DBI values for the PI483463 dataset. The value achieved by SRSIO-FCM and SLFCM is much better for the 60d-SPF than 6d-SCPSF. As we can see, for 6d-SCPSF, DBI values are higher than 60d-SPF when clustered using SRSIO-FCM and SLFCM on almost all the clusters. On the other hand, the existing methods (Bandyopadhyay and Mansoori) have extremely higher values than those proposed.  Moreover, SRSIO-FCM attained a low value in cluster 5 for 60d-SPF.  

  Table \ref{dbi-W05} shows the DBI values for the W05 dataset. The value achieved by SRSIO-FCM and SLFCM is much better for the 60d-SPF than 6d-SCPSF. As we can see, for 6d-SCPSF, DBI values are higher than 60d-SPF when clustered using SRSIO-FCM and SLFCM on almost all the clusters. On the other hand, the existing methods (Bandyopadhyay and Mansoori) have extremely higher values than those proposed.  Moreover, SRSIO-FCM and SLFCM attained a low value in cluster 5 for 60d-SPF.  
  
Finally, we can conclude that 60d-SPF performs better than the proposed 6d-SCPSF extraction and existing methods (Bandyopadhyay and Mansoori) when applied to SRSIO-FCM and SLFCM algorithms. The performance is reported in SI and DBI values for Lee, Williams82, PI483463, and W05 soybean protein datasets.

\section{Conclusion and Future Work} \label{sec5}
 In this paper, two scalable feature extraction approaches, i.e., 60d-SPF and 6d-SCPSF, have been proposed to extract a numerical feature vector from huge protein sequences using the Apache Spark cluster. After that, preprocessed numerical feature vectors are applied to the scalable fuzzy clustering algorithms. In this case, we have used the SRSIO-FCM and SLFCM algorithms as the scalable fuzzy clustering approaches to cluster huge soybean protein datasets. One of the most significant characteristics of the proposed scalable feature extraction approach is that it takes raw protein sequences of variable length as input and produces fixed-length numeric feature vectors as an output. Thus, both the algorithms are scalable and can handle huge numbers of protein sequences. However, when obtained feature vectors are applied to scalable fuzzy clustering algorithms, the 60d-SPF extraction method excels over the 6d-SCPSF and existing feature extraction approaches. One distinctive characteristic of the proposed 60d-SPF approach is that it computes pseudo amino acid composition parameters on the Apache Spark cluster to consider all possible position-specific variations of amino acids in a protein sequence. We focused on the exact evaluation of both the proposed feature extraction methods applied to the SRSIO-FCM and SLFCM clustering algorithms. The performance was investigated on distinct soybean protein datasets. The investigated results demonstrate the potential benefits of adopting our feature extraction technique for massive protein datasets. In the future, we are interested in applying and analyzing the performance of the proposed scalable feature extraction techniques on other plant species like rice and wheat data.

\begin{acknowledgements}
This work is supported by National Supercomputing Mission, HPC Applications Development Funded Research Project by DST in collaboration with the Ministry of Electronics and Information Technology (MeiTY).
\end{acknowledgements}

\section*{Compliance with ethical standards}
\textbf{Conflict of interest} All authors declare that there are no conflicts of interests.

%

\bibliographystyle{spbasicunsort}      
\nocite{*}

\bibliography{ref}   

%
%

\end{document}